\documentclass[12pt,preprint]{aastex}

\usepackage{color}
\usepackage{graphicx}
\usepackage{amsmath}
\usepackage{natbib}

\shorttitle{Prompt EM Luminosity}
\shortauthors{Krolik}

\begin{document}

\title{Estimating the Prompt Electromagnetic Luminosity of a Black Hole Merger}

\author{Julian H.\ Krolik}
\affil{Department of Physics and Astronomy,
Johns Hopkins University\\
Baltimore, MD 21218}
\email{jhk@pha.jhu.edu}

\begin{abstract}

    Although recent work in numerical relativity has made tremendous strides in
quantifying the gravitational wave luminosity of black hole mergers, very little
is known about the electromagnetic luminosity that might occur in immediate
conjunction with these events.  We show that whenever the heat deposited in the
gas near a pair of merging black holes is proportional to its total mass, and
the surface density of the gas in the immediate vicinity is greater than the
(quite small) amount necessary to make it optically thick, the characteristic
scale of the luminosity emitted in direct association with the merger is
the Eddington luminosity {\it independent} of the gas mass.  The
duration of the photon signal is proportional to the gas mass, and is generally
rather longer than the merger event.  At somewhat larger distances,
dissipation associated with realigning the gas orbits to the new spin orientation
of the black hole can supplement dissipation of the energy gained from orbital
adjustment to the mass lost in gravitational radiation; these two heat sources
can combine to augment the electromagnetic radiation over longer timescales.

\end{abstract}

\section{Introduction}

    The merger of two supermassive black holes has been a topic of lively
astrophysical speculation for many years \citep{BBR80}.
Recent developments in galaxy formation theory have made the prospect more
plausible and suggest an environment for such events: the centers of galaxies
that underwent major mergers a few hundred million years in the past
\citep{HaehKauff02,Volont03}.  Mergers may be particularly likely when
the galaxy contains a relatively rich supply of interstellar
gas, which may help binary black holes overcome the ``last
parsec problem" and approach each other close enough for gravitational wave
emission to compress the orbit to merger within a Hubble time
\citep{GR00,AN02,Kazantz04,Escala05,Cuadra09,Dotti09}
(but for a contrary view see \cite{Lodato09}).  The
presence of sizable quantities of interstellar gas in the parsec-scale
environment then raises the question of how much gas might find itself
even closer to the pair at their moment of merger.

     This is a subject of great uncertainty.  It has been argued, for example,
that there should be little gas closer to the merging pair than
$\sim 100$--$1000r_g$ ($r_g \equiv GM/c^2$, where $M$ is the total mass of the
system) because eventually the timescale
for shrinkage of the binary orbit by gravitational wave radiation becomes
shorter than the timescale for mass inflow due to internally generated
fluid stresses \citep{MP05}.  On the other hand, the mass of such a circumbinary
disk might be as large as $\sim 100 M_{\odot}$ or more
\citep{MP05,AN02,Rossi09,Corrales09};
if even $1 M_{\odot}$ were close enough to the merging black holes to
be given heat equal to $1\%$ of its rest mass,
the total energy---$\sim 10^{50}$~erg---might well be large enough to produce
observable radiation.  It is therefore a worthwhile
exercise to estimate what sort of light might be generated if even a small
fraction of the surrounding gas were able to make its way in close to the merging
black holes.

Because the amount of mass near the merging black holes is so difficult to
estimate at present, the plan of this paper is to explore prompt electromagnetic
radiation in a way that is scaled to whatever gas mass is there.  Thus, we will first
estimate the heat per unit mass that might be deposited in this gas, then,
in order to find the luminosity, estimate the timescale on which the energy is
radiated.  Next, the more model-dependent subject of the spectrum will
be broached.  Lastly, having seen how the light emitted depends on gas mass,
we will discuss the issues related to whether an ``interesting" amount of
mass may be present.  In order to avoid additional complications, we will ignore any
luminosity due to accretion through the circumbinary disk.

\section{Efficiency of Energy Deposition}

    Let us begin, then, with the supposition that immediately before a merger
of two supermassive black holes there is at least some gas orbiting over a range of
distances not too far from the system's center of mass.  To discuss the effect of
the merger on this relatively nearby gas, it is useful to distinguish two
regions: the inner gas ($r/r_g \lesssim 10$) and gas farther away
($10 < r/r_g < 10^3$).  These regions are distinguished both by the magnitude of
the heating they are likely to experience and by the time at which it occurs.

Because the inner gas is in the ``near-field" regime, its gravitational environment
during the merger is better described as a nonlinear time-dependent distortion of
spacetime, rather than a passage of gravitational waves.  The amplitude and
extent of the distortions are, in some sense, proportional to the binary
mass ratio, reaching a maximum when the two black holes have similar mass.
Strong shearing can deflect fluid orbits, provoking shocks, and can also
stretch magnetic field lines, driving MHD waves whose energy can ultimately
be dissipated.  Because the characteristic timescale of the spacetime variability
is $\sim O(10)r_g/c$ \citep{BCP07}, while the dynamical timescale of particle
orbits in this region is only a little longer ($\sim (r_g/c)(r/r_g)^{3/2}$), the
fluctuations can be very efficient in transferring energy to the gas (cf. the
test-particle calculations of \citet{vanMeter09}).  The same near-coincidence of
merger timescale with the gas's dynamical timescale means that the dissipation
timescale (via shocks, etc. which develop on a dynamical timescale) should
also be comparable to the merger
duration.\footnote{\cite{KocLoeb08} suggested that the dissipation should be
related to shear in the way characteristic of steady-state accretion disks.
However, in the present case, the shear varies much more rapidly than the
saturation time for the MHD turbulence whose dissipation is relevant to
steady-state disks.  The classical relationship is therefore unlikely to apply.}
Unfortunately, without detailed general relativistic MHD calculations, it is
very difficult to make a quantitative prediction of just how much energy might
thus be given to the gas.

Nonetheless, to the extent that gravitational dynamics dominate, the Equivalence
Principle suggests that the energy left in
the gas should be proportional to its mass.  It is therefore convenient
to scale the gas heating in familiar rest-mass efficiency terms.  For a fiducial
value, one might imagine that this efficiency $\epsilon$ could be as large as
$\sim 0.1$ for matter subjected to a truly nonlinear dynamical spacetime.
Gravitational shear acting on magnetic
fields may supplement this energy deposition \citep{MK04,Duez05,Pal09}.

     At greater distances, the perturbations to spacetime are much smaller and
the contrast between the wave frequency and the dynamical frequency is much greater.
In addition, there can be a considerable delay between the time of the merger and
the time at which the principal heating occurs.  Three mechanisms can cause heating
at these larger distances, one resulting from the sudden loss of mass from the
merged black hole due to its emission of gravitational waves \citep{BodePh07},
another due to its sudden loss of angular momentum, and a third the result of
the merged black hole's recoil as a result of asymmetric gravitational
wave radiation \citep{LFH08}.  We stress, however, that the radial scales whose
radiation is under consideration here are {\it not} those responsible for
the longer-term afterglow that has been the focus of prior work
\citep{MP05,LFH08,SB08,SK08,Rossi09,Corrales09}; the afterglow is due to
heating of mass in the circumbinary disk proper, at the relatively large
radii (at least $\sim 100 r_g$: \cite{MP05}) where conventional accretion dynamics
were able to bring it during
the time when gravitational wave evolution of the binary was faster than
the typical inflow rate.  Although the relevant heating mechanisms in the radial
range considered here are very similar
to those acting at larger radii, the focus of this work is on matter interior to
the disk proper, where any gas present arrived as the result of angular momentum
loss faster than that acting on the bulk of the circumbinary disk.  Note, too,
that because we restrict our attention to $r/r_g < 10^3$, all the gas remains
bound to the merged black hole for even the largest of recoil velocities.

      As a result of the mass lost by the black hole in gravitational wave radiation,
the binding energy of orbiting matter is immediately reduced by
$(\Delta M/M)(r/r_g)^{-1}$, where the fractional black hole mass loss
$\Delta M/M \sim 1$--$10\%$ \citep{Berti07,Rezz08,LCZ09} with the exact
number depending on the mass ratio and spins
of the merging pair.  Relative to the dynamical timescale, this change in energy
occurs almost instantaneously because the merger duration is so much shorter than
an orbital period when $r \gg r_g$.  However, any heating due to this change in
energy is delayed by $\sim (r_g/c)(r/r_g)^{3/2}(\Delta M/M)^{-1}$ because the
eccentricity
induced in the orbits is only $\sim \Delta M/M$ \citep{SK08,ONeill09}.  In
addition, it is possible that only a fraction $\Delta M/M$ of the orbital energy
gained is dissipated, as most of the
energy may be used simply to expand the gas orbits \citep{ONeill09,Corrales09,Rossi09}.

There can also be dissipation due to the sudden change in angular momentum.
Torques driven by the binary's quadrupolar mass distribution acting on matter
surrounding the pre-merger binary will cause any obliquely-orbiting gas to precess
around the direction of the binary's total angular momentum (at large separation,
the total angular momentum is almost exactly the orbital angular momentum);
dissipation between intersecting
fluid orbits should then lead to alignment of the gas's orbital angular
momentum with the binary's angular momentum \citep{LubowOg00}.  If there is any
misalignment between either of the two spin directions
and the orbital angular momentum, the angular momentum of the merged black hole could
be in a different direction from the original total angular momentum
\citep{Schn04,Bogdan07,Rezz08}.  After the merger, Lense-Thirring torques will act in a
fashion closely analogous to the Newtonian torques acting during the binary phase,
and dissipation should then reorient orbiting gas into the new equatorial plane
\citep{BP75}.  Because the kinetic energy of motion out of the new equatorial plane
is a fraction $\sim \sin^2(\Delta\theta)$
of the orbital energy for misalignment angle $\Delta\theta$, the amount of energy
dissipated can be an order unity fraction of the orbital energy.
The delay from the time of merger to when this mechanism acts will be of
order the precession time, $\sim (r/r_g)^3 (r_g/c)$, which is rather longer than the
delay before dissipating the orbital energy gained from mass-loss wherever
$r/r_g > (\Delta M/M)^{-2/3}$.  However,
the degree of misalignment is highly uncertain and may depend strongly on
details of the environment; for example, accretion during the inspiral may align
both black hole spins with the orbital angular momentum, eliminating a change
in angular momentum direction as a result of the merger \citep{Bogdan07}.
Close alignment of both spins with the orbital angular
momentum will, however, have the compensating effect of producing an especially
large $\Delta M/M$ \citep{LCZ09}.  Because both the mass-loss and the
Lense-Thirring effects scale with the local binding energy, we will combine them,
writing their efficiency as
$\epsilon \simeq \eta (r/r_g)^{-1}$, where we expect $\eta \lesssim 0.01$.

   Lastly, response to the black hole recoil adds an energy per unit mass to
the disk matter $\sim v_{\rm recoil}^2$.  Even if only a fraction $\Delta M/M$ of
the mass-loss energy leads to heating, the recoil energy becomes larger than
the heating due to mass-loss only at radii
\begin{equation}
r > \left(\frac{\Delta M/M}{v_{\rm recoil}/c}\right)^2 
       \simeq 10^3 \left(\frac{\Delta M/M}{0.03}\right)^2 
           \left(\frac{v_{\rm recoil}}{300\hbox{~km~s$^{-1}$}}\right)^{-2} ,
\end{equation}
\citep{SK08,Rossi09}, so we neglect it in the estimates presented here.

\section{Cooling Time and Luminosity}

    Whatever heat is deposited in the gas, the rate at which this energy is carried
away by radiation is determined by how rapidly electrons can generate photons and
then by how rapidly those photons can make their way outward through the
opacity presented by the material itself.  We will consider the first
issue later (\S~\ref{sec:spectrum}) when we discuss the complications of
estimating the radiating gas's temperature.  For the time being we will assume
that photon diffusion is the slower of the two processes.  Unless $\epsilon$
is extremely small, the gas is likely to be so thoroughly ionized that electron
scattering dominates its opacity, so the optical depth is simply proportional to the
surface density.

To be optically thin, the surface density $\Sigma$ must then be very small:
$\Sigma < 3$~gm~cm$^{-2} \simeq 3 \times 10^{-9} M_7^2 (M_{\odot}/r_g^2)$, where
$M_7$ is the total black hole mass in units of $10^7 M_{\odot}$.  In this case,
if the gas is able to convert heat into photons at least as fast as the
heat is delivered, the luminosity per logarithmic radius is
the ratio of energy deposited to the time in which it is dissipated,
roughly the dynamical time:
\begin{equation}
\frac{dL}{d\ln r} \simeq 2.5 \times 10^{44} \epsilon(r) \Sigma (r/r_g)^{1/2} M_7
\hbox{~erg~s$^{-1}$}
\end{equation}
for $\Sigma$ in gm~cm$^{-2}$.
The duration of such a flare should be only $\sim 50 (r/r_g)^{3/2}M_7$~s.

    A more interesting regime is presented by the case of an optically thick
region.  Under optically thick conditions, if the vertical scaleheight of the
disk is $h$, the cooling time
$t_{\rm cool} \sim \tau h/c \sim \tau (h/r)(r/r_g)(r_g/c)$, where the optical
depth from the midplane outward is $\tau = \kappa \Sigma/2$ for opacity per unit
mass $\kappa$.  Because the duration
of the merger event is only $\sim O(10) r_g/c$, unless the disk is able to stay
very geometrically thin (which the following argument will demonstrate is unlikely),
the heat given the gas during the
merger will be radiated over a time much longer than the merger event proper.

    The disk thickness is controlled by two elements: the initial heat content of the
disk and the heat deposited as a result of the merger.  We will neglect the
former partly because it is so uncertain, partly because it seems plausible that
it will be outweighed by the latter, and partly because this represents in
some respects a conservative assumption.

Suppose, then, to begin that the optical depth is large enough to make the cooling time
longer than the orbital time (i.e., $\tau (h/r) > (r/r_g)^{1/2}$).  There will then
be time for the gas to achieve a dynamical equilibrium (if one is possible).
Radiation pressure will likely dominate gas pressure because the photon escape time
is much longer than the photon radiation time (the reasoning behind this
assertion is discussed in \S~\ref{sec:spectrum}).  Put another way, nearly
all the thermal energy density initially given the matter is rapidly
converted into photons; when the cooling time is longer than the dynamical time,
they are still present for many dynamical times.  Consequently, their pressure
becomes much greater than the gas pressure.

The force exerted by the (slowly) diffusing photons is proportional to their
flux times the opacity; because the flux is
the energy per unit area per cooling time, we can determine $h$ by matching the
vertical radiation force to the vertical component of gravity.
Here, in order to obtain a rough estimate of the disk thickness, we suppose
that $h \ll r$ and that Newtonian gravity applies:
\begin{equation}
\frac{\kappa\epsilon \Sigma c^2}{2 \tau h} \simeq h \Omega^2,
\end{equation}
which leads to
\begin{equation}
h/r \simeq \epsilon^{1/2} (r/r_g)^{1/2}.
\end{equation}
In other words, the geometric profile of optically thick gas immediately post-merger
depends only on the heating efficiency $\epsilon$ (if it were optically thin,
$h/r \simeq (\tau \epsilon r/r_g)^{1/2}$).  Moreover, if this equilibrium is
achieved, the criterion that the cooling time exceed the dynamical time is
easily achieved in optically thick disks if $\epsilon$ is not too small, for
all that is required is $\tau > \epsilon^{-1/2}$.

Close to the black hole,
where $\epsilon$ may be as much as $\sim O(0.1)$, the disk may be almost
spherical.  It is imaginable that $\epsilon$ is
so large that no hydrostatic equilibrium is possible (i.e., $h/r > 1$); in that
case, the radiation flux would drive the gas away from the black hole merger
remnant.  For the purposes of this order-of-magnitude treatment, we ignore that
possibility; we also ignore further order-unity corrections that might result from
time-dependent photon diffusion effects.  Farther from the black hole, where
$\epsilon \sim \eta (r/r_g)^{-1}$, the disk should be thinner: $h/r \sim \eta^{1/2}$.
In these more distant regions, cancellation of the radial scalings leaves the
thickness to be determined by the details of local heat dissipation (i.e.,
the effectiveness of dissipating the energy gain due to mass-loss and
whatever disk re-orientation takes place).

      With the disk thickness estimated, the cooling time immediately follows:
\begin{equation}
t_{\rm cool} = \tau h/c \simeq 50 \tau \epsilon^{1/2} (r/r_g)^{3/2} M_7\hbox{~s}.
\end{equation}
Equivalently, it is $\simeq (\tau/2\pi)\epsilon^{1/2}$ orbital periods.

Although the radiating timescale is proportional to the surface density,
the luminosity is {\it independent} of it so long as the heating time is shorter
than the cooling time.  This is simply because
the energy to be radiated is proportional to the surface density,
while the cooling time is likewise:
\begin{equation}\label{eq:lum}
\frac{dL}{d\ln r} \simeq 1.5 \times 10^{45} \epsilon^{1/2} (r/r_g)^{1/2}
                      M_7 \hbox{~erg~s$^{-1}$} = \epsilon^{1/2} (r/r_g)^{1/2} L_E .
\end{equation}
The luminosity scale is the Eddington luminosity because the time to cross the
heated (and inflated) radiating region is proportional to how well the radiation
flux can resist gravity.  How close the luminosity approaches to Eddington is
determined by the efficiency.

Farther from the black hole, the efficiency $\epsilon = \eta (r/r_g)^{-1}$,
so if $t_{\rm cool} > t_{\rm heat}$ and $\tau > 1$,
\begin{equation}
\frac{dL}{d\ln r} \simeq \eta^{1/2} L_E.
\end{equation}
That is, the luminosity from regions where $r/r_g \gg 1$ should scale with the
Eddington luminosity, but may be a fairly small fraction of it.  

However, in these more distant regions (i.e., $10 \lesssim r/r_g < 10^3$),
the heating time can be so extended that it might be
longer than the cooling time, particularly if the optical depth is not very large:
\begin{equation}
\frac{t_{\rm cool}}{t_{\rm heat}} \sim \tau \eta^{1/2}\begin{cases}
(r/r_g)^{-1/2} & \text{mass-loss}\\
(r/r_g)^{-2} & \text{Lense-Thirring}\end{cases}.
\end{equation}
When the dynamical response of the disk is so slow that it exceeds the cooling time,
the luminosity becomes
\begin{equation}
\frac{dL}{d\ln r} = \eta \tau L_E\begin{cases}
(r/r_g)^{-1/2} & \text{mass-loss} \\
(r/r_g)^{-2} & \text{Lense-Thirring}\end{cases}.
\end{equation}

To find the total observed luminosity, we must assemble the luminosity from different
regions, making proper allowance for their different time-dependences.  Initially,
the radiative output will be dominated by the inner radii, so that
$L \simeq (\epsilon r_{\rm in}/r_g)^{1/2}L_E$, where $r_{\rm in}$ is the scale of
the region subject to the truly dynamical spacetime.  After a time
$50 \tau \epsilon^{1/2} (r_{\rm in}/r_g)^{3/2} M_7$~s, this light decays, to
be replaced over longer timescales by the signals due to mass-loss and the
Lense-Thirring reorientation: $\sim 50 (r/r_g)^{3/2} M_7 (\Delta M/M)^{-1}$~s for
the former, $\sim 50 (r/r_g)^3 M_7$~s for the latter.  As estimated above, the
luminosity from these regions at somewhat larger radius is likely rather less
than the luminosity issuing from the innermost region.

\section{Temperature and Spectrum}\label{sec:spectrum}

Prediction of the output spectrum is much more model-dependent.  If there is enough
absorptive opacity to thermalize the radiation, its characteristic temperature would be
\begin{equation}
T \sim 1 \times 10^6 \epsilon^{1/8} (r/r_g)^{-3/8} M_7^{-1/4} \hbox{~K},
\end{equation}
similarly universal, decreasing only slowly with increasing black hole mass.

Whether thermalization can be achieved, however, may be sensitive to
conditions.  The effectiveness of free-free opacity can be enhanced by the
additional path-length to escape imposed upon the photons by scattering.
At the temperature just estimated, the effective optical depth (i.e., the
geometric mean of the free-free and Thomson optical depths) for photons
near the thermal peak is
\begin{equation}
\tau_{eff} \simeq 3 \times 10^{-4} \tau^{3/2} \epsilon^{-15/32}
                     (r/r_g)^{-3/32} M_7^{-1/16}.
\end{equation}
Thus, if free-free is the only absorption mechanism,
$\tau \gtrsim 100 (\epsilon/0.1)^{5/16}$ is required to achieve thermalization.
On the other hand, where the temperature is $\sim 10^5$~K or less, atomic transitions
substantially enhance the absorptive
opacity, making thermalization much more thorough.
Thus, the detailed character of the emitted spectrum could vary considerably from
case to case.

     With an estimate of the temperature, we can now estimate the timescale on
which electrons radiate most of the heat given the gas.  Considering only
free-free emission, it is
\begin{equation}
t_{\rm rad} \simeq 60 \epsilon^{1/2} (r/r_g)^{3/2} M_7 \tau^{-1} T_5^{1/2}\hbox{~s}.
\end{equation}
In other words, $t_{\rm rad}/t_{\rm cool} \simeq 1.2 \tau^{-2} T_5^{1/2}$, so
that thermal balance at a temperature $\lesssim 10^5$~K is a self-consistent
condition for an optically thick region.

However, unless $\epsilon$ is exceedingly small, the gas's temperature immediately
upon being heated will be far higher than $10^5$~K.  If the initial shocks driven
by the dynamical spacetime have speeds comparable to the orbital speed, the
post-shock electron energies will more likely be well in excess of 1~MeV, and the
characteristic radiation rate will be much slower.  In this initial high-temperature
state, the free-free radiation time is
\begin{equation}
t_{\rm rad} \simeq 1 \times 10^4 \epsilon^{1/2} (r/r_g) M_7 \tau^{-1}
    \left( \ln \Theta\right)^{-1}\hbox{~s},
\end{equation}
where $\Theta$ is the electron energy in rest-mass units (assumed to be $> 1$
in this expression), and we have
estimated the disk aspect ratio $h/r \simeq \epsilon^{1/2}$ (by assumption,
at this stage radiation pressure is not yet important).  Comparing this estimate
of $t_{\rm rad}$ to $t_{\rm cool}$, we find that
\begin{equation}\label{eq:freefreerad}
\frac{t_{\rm rad}}{t_{\rm cool}} = 200 \tau^{-2} \left( \ln \Theta\right)^{-1},
\end{equation}
entirely independent of $h/r$ (as well as $r/r_g$ and $M$).  Thus, 
if free-free radiation is the only photon-creation mechanism, optical depths
$\gtrsim O(10)$ would be required in order to create photons carrying most of
the heat in a time shorter than it takes for those photons to escape.

There are, however, other processes that can also likely contribute.  Suppose,
for example, that the magnetic field energy density is a fraction $q$ of the
plasma pressure.  Synchrotron radiation would then cool the gas on a timescale
\begin{equation}
t_{\rm rad} \sim 50 q^{-1}\tau^{-1}\Theta^{-2} \epsilon^{1/2}(r/r_g)M_7\hbox{~s}
\end{equation}
wherever $\Theta > 1$.  Relative to the cooling time, this photon production
timescale is
\begin{equation}
\frac{t_{\rm rad}}{t_{\rm cool}} \sim q^{-1}\tau^{-2}\Theta^{-2}.
\end{equation}
Inverse Compton radiation would be equally effective if the energy density
in photons of energy lower than $\sim \Theta m_e c^2$ is that same fraction
$q$ of the plasma pressure.  As shown by equation~\ref{eq:freefreerad},
relativistic electron free-free radiation is able to radiate at least
$\sim 10^{-2}$ of the heat during a photon diffusion time if $\tau > 1$;
we therefore expect the $q$ in photons to be at least this large.  Thus,
even a small initial cooling by free-free radiation, particularly when
supplemented by a modest magnetic field, should provide enough seeds for
inverse Compton cooling to allow the gas to radiate the great majority of
its heat content in a time shorter than the photon diffusion time.  All
that is required is $\tau > 10 (q/0.01)^{-1/2}\Theta^{-1}$.

Even if the majority of the dissipated energy is given to the ions, so that
their temperature is larger than the electrons', rapid radiation is still
likely to occur.  The ratio between the timescale for thermal coupling between
ions and electrons by Coulomb collisions and the photon diffusion time is
\begin{equation}
\frac{t_{\rm ep}}{t_{\rm cool}} \simeq 60 \tau^{-2} \Theta^{3/2}.
\end{equation}
Thus, provided $\tau > 8 \Theta^{3/4}$, the plasma should achieve a one-temperature
state more rapidly than the photons can leave.


\section{What is $\Sigma(r)$?}

Lastly, we turn to the hardest question to answer at this stage:
how much gas there should be as a function of radius, here parameterized
as $\tau(r)$.  Particularly in the inner region, it would take very little gas
to create a large optical depth: even integrated out to $r/r_g=100$, a disk
with constant $\tau = 100$ would require only $10^{-4}M_7^2 M_{\odot}$.

Even though there is no reason to think the disk is anywhere
near a conventional state of inflow equilibrium, one could use the optical
depth of such a disk as a standard of comparison.
The thermodynamics of equilibrium disks creates a characteristic scale
for the surface density: the maximum
at which thermal equilibrium can be achieved.
One of the predictions of the \cite{SS73} model is that in a
steady-state disk in which the vertically-integrated $r$-$\phi$ stress is $\alpha$
times the vertically-integrated total pressure, the accretion rate at any particular
radius increases as the
surface density increases, but only up to a point.  Larger surface density (and
accretion rate) lead to a larger ratio of radiation to gas pressure.  If
radiation pressure exceeds gas pressure, increasing accretion rate can only
be accommodated by a {\it decreasing} surface density.  In other words,
there is a {\it maximum} possible surface density.  Although recent work on
explicit simulation of disk thermodynamics under the influence of MHD turbulence
driven by the magneto-rotational instability has shown that this phenomenological
model's prediction about the thermal stability of disks is wrong \citep{Hirose09a},
they also show that vertically-integrated disk properties averaged over times
long compared to a thermal time {\it do} match those predicted by the $\alpha$ model
\citep{Hirose09b}: when radiation pressure dominates, the surface density and
accretion rate are inversely related.  The Thomson optical depth corresponding to
this maximum surface density is
\begin{equation}
\tau \simeq 2.5 \times 10^4 (\alpha/0.1)^{-7/8} M_7^{1/8}(r/10r_g)^{3/16},
\end{equation}
where we have scaled the stress/pressure ratio to 0.1.  Close to the black
hole, it occurs at a comparatively low accretion rate in Eddington units:
$$\dot m \simeq 1.4 \times 10^{-3} (\alpha/0.1)^{-1/8} (r/10r_g)^{21/16} M_{7}^{-1/8}.$$
Such a state might be consistent with a gas supply rate at large radius capable
of feeding an AGN (i.e., $\dot m \sim 0.1$), but reduced two orders of magnitude
by the effects of binary torques and the inability of internal stresses in the
disk to drive its inner edge inward as fast as gravitational wave emission compresses
the black hole binary.  It is significant in this respect that even such a
strong suppression of accretion still yields an inner disk optical depth that
is quite large.  A smaller accretion rate would produce a smaller optical depth,
but only $\propto \dot m^{3/5}$, when gas pressure dominates and the disk remains
radiative.

These estimates can also serve as a springboard to gain a sense of what might
occur in states of inflow non-equilibrium.   For example,
if gas accretes at larger radii but is held back by binary orbital torques at
radii several times the binary separation, its surface density at the point
where it is held back should be larger than
what would be expected on the basis of equilibrium inflow at the accretion
rate farther out.  In such a case, although the preceding estimates
might be reasonable at larger radii, the optical depth could be substantially
enhanced closer to the merging binary.

Another uncertainty is presented by the question of whether the optical depth
in a given region remains the same over the entire radiating period.  In the
inner disk, for example, there could be significant radial motions, both inward
or outward, engendered directly by the dynamical spacetime during the merger event.
Because the cooling time in the inner disk is rather longer than an orbital period,
there might be time for the magneto-rotational instability to stir MHD turbulence
that could drive accretion and restore some of the merger heat carried off
by radiation.  If the optical depth is relatively small, so that
$\tau \epsilon^{3/2} \alpha (r_{\rm in}/r_g) < 1$ and
the accretion time in the inner disk is longer than the cooling time, the luminosity
would gradually taper off as the accretion luminosity extends the bright period, but
disk cooling causes the accretion rate to diminish.  Alternatively, if the optical
depth is larger, the inflow time would
actually be shorter than the cooling time.  In this case, the luminosity would
be greater than earlier estimated, but the lifetime of bright emission from the
inner disk would be reduced to the inflow time.   In either case, the total energy
released due to accretion could increase the total emission by a factor of order
unity or more because the radiative efficiency of accretion near a black hole
is generically $\sim O(0.1)$.

\section{Summary}

To summarize, when a pair of supermassive black holes merge, provided only that
the gas very close to the merging pair has at least a small electron scattering
optical depth, we expect the prompt EM signal to likely have a luminosity comparable to
the Eddington luminosity of the merged system,
$\sim 10^{45}\epsilon^{1/2} M_7$~erg~s$^{-1}$.  The duration of this bright
phase is proportional to the mass of gas at $\sim 10r_g$ from the merged black
hole, a quantity that is at present
extremely difficult to estimate.  If there is only the very small
amount of gas necessary to be optically thick, the duration would be only
$\sim 10M_7$~min; on the other hand, quantities several orders of magnitude
larger are well within the range of plausibility.  Gas at somewhat greater
distance ($10 < r/r_g < 10^3$) should continue to radiate at a level possibly
as high as $\sim 0.1 L_E$, but for a longer time.

Because gravitational wave observatories like the {\it Laser Interferometric Space
Antenna} (LISA) are expected to give approximate source localization
days or even weeks in advance of merger \citep{LH09}, one could hope for
synergistic observing campaigns that might catch the entire EM signal.
Alternatively, it may be feasible to use EM surveys with large solid-angle
coverage to
search for source flaring having the characteristics predicted here in order
to identify candidate black hole merger systems before any gravitational
wave detectors are ready.

     The spectrum of this light is, at present, much more difficult than its
luminosity to predict with any degree of confidence.  It may peak in the
ultraviolet, but if it does, it is likely to
be rather bluer than the familiar UV-peaking spectra of AGN (if the spectrum
does peak in the UV, extinction in the host galaxy may obscure some number
events).  While the
inner region still shines, its luminosity will likely be greater than
that from greater radii, making the spectrum (to the degree it is
thermalized) closer to that of a single-temperature system.  If it
is only incompletely thermalized, a still harder spectrum might be
expected.  At later times, when the outer disk dominates, the temperature
corresponding to a given radius is $\propto \eta^{1/8} (r/r_g)^{-1/2} M_7^{1/4}$.
Compared to the temperature profile of an accretion disk in inflow equilibrium
($T \propto (r/r_g)^{-3/4}$), this is a slower decline outward, suggesting
a spectrum that might be softer than during the initial post-merger phase,
but still harder than that of typical AGN.  In addition, over time its
high frequency cut-off will move to lower frequencies.  Because our understanding
of what determines the spectra of accretion disks around black holes is far
less solid than our understanding of their gross energetics (consider, for
example, the still-mysterious fact that ordinary AGN radiate a significant
fraction of their bolometric luminosities in hard X-rays), these spectral
predictions are far shakier than the predicted bolometric luminosity scale.

After the merger heat has been radiated, the disk should revert to more normal
accretion behavior and, beginning at the smallest radii, should display
standard AGN properties \citep{MP05}.
Unfortunately, the pace of fading as a function of radius depends on the run of
surface density with radius, which at the moment is highly uncertain.  It
is therefore far beyond the scope of even this speculative paper to guess
how rapidly that may occur.

\acknowledgements
This work was partially supported by NSF grant AST-0507455 and NASA Grant NNG06GI68G.
I am grateful to Omer Blaes, John Hawley, and Jeremy Schnittman for comments on
the manuscript, and to both Manuela Campanelli and Jeremy Schnittman
for stimulating my interest in this subject.

\bibliographystyle{apj}
\bibliography{../references}

\end{document}